\documentclass[aps,pre,twocolumn]{revtex4}
\usepackage{graphicx}
\usepackage{amsmath}
\usepackage{amssymb}
\usepackage{bm}
\usepackage{pifont}
\usepackage{color}

\begin{document}

\title{Polarized Ukraine 2014: Opinion and Territorial Split Demonstrated with the Bounded Confidence XY Model, Parameterized by Twitter Data}

\author{Maksym Romenskyy}
\affiliation{Department of Life Sciences, Imperial College London, London SW7 2AZ, UK}
\affiliation{Department of Mathematics, Uppsala University, Box 480, Uppsala 75106, Sweden}
\author{Viktoria Spaiser}
\affiliation{School of Politics and International Studies, University of Leeds, Leeds LS2 9JT, UK}
\author{Thomas Ihle}
\affiliation{Institute of Physics, University of Greifswald, Felix-Hausdorff-Str. 6, Greifswald 17489, Germany}
\author{Vladimir Lobaskin}
\affiliation{School of Physics, University College Dublin, Belfield, Dublin 4, Ireland}

\date{\today}
\begin{abstract}
Multiple countries have recently experienced extreme political polarization, which in some cases led to escalation of hate crime, violence and political instability. Beside the much discussed presidential elections in the United States and France,  Britain's Brexit vote and Turkish constitutional referendum, showed signs of extreme polarization. Among the countries affected, Ukraine faced some of the gravest consequences. In an attempt to understand the mechanisms of these phenomena, we here combine social media analysis with agent-based modeling of opinion dynamics, targeting Ukraine's crisis of 2014. We use Twitter data to quantify changes in the opinion divide and parameterize an extended Bounded-Confidence XY Model, which provides a spatiotemporal description of the polarization dynamics. We demonstrate that the level of emotional intensity is a major driving force for polarization that can lead to a spontaneous onset of collective behavior at a certain degree of homophily and conformity. We find that the critical level of emotional intensity corresponds to a polarization transition, marked by a sudden increase in the degree of involvement and in the opinion bimodality.
\end{abstract}
\maketitle

\section{Introduction}

\begin{figure*}
\begin{center}
\includegraphics[scale=0.5]{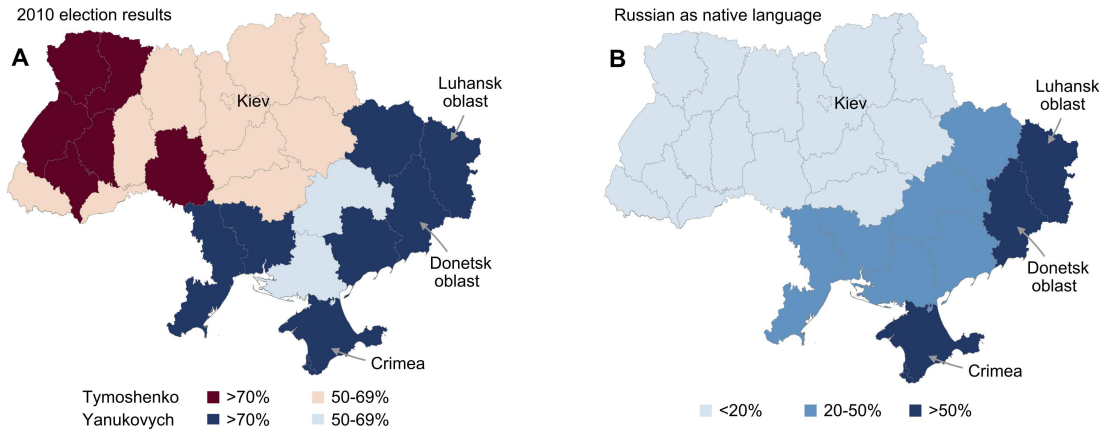}
\end{center}
\caption{(A) Ukraine's political divide in 2010 elections \cite{BBC}. The majority of the voters in the eastern and southern regions of Ukraine supported a pro-eastern candidate, Viktor Yanukovych. (B) Ukraine's linguistic divide according to national census 2001 \cite{cencus}. In the three regions most affected by Ukraine crisis in 2014, Luhansk oblast, Donetsk oblast and Crimea, Russian is a native language for more than 50\% of the population.}
\label{bbc}
\end{figure*}

Ukraine represents a bright example of a nearly evenly split society with two opposing camps, where the East/South gravitates towards Russia while the West/North towards European neighbors \cite{Belletal14}. The overall political vector in the country sways between political parties and leaders that on the one side seek closer ties to the West and in particular Europe and on the other hand to the East and in particular Russia. The Orange Revolution in 2004 brought pro-western politicians to power, however, in the 2010 elections a pro-eastern politician, Viktor Yanukovych, was elected for president, not least because of major support in the eastern regions of Ukraine (see Fig. \ref{bbc}). In November 2013, after Yanukovych failed to sign a political association and free trade agreement with the European Union, protests in Ukraine erupted. The initially peaceful rallies became violent in January 2014 after the government passed laws to suppress the protests. In February 2014, the violence escalated, which led to the removal of Yanukovych from office by the parliament. Meanwhile a separatist and anti-interim-government movement rose with the support of Russia in eastern and southern parts of Ukraine, and ignited a military conflict. Crimea was annexed by the Russian Federation after a referendum that was denounced internationally as illegitimate and illegal. Later in 2014, the crisis resulted in further territorial separation with over 2.6 million internally displaced persons and refugees and a formation of self-proclaimed states in Donetsk and Luhansk \cite{EuropeanCommission16}. These events escalated the polarization in the country that has grown over the years. As the two political sides became more extreme in their views in the course of the events a dialogue and therefore a peaceful solution has become increasingly difficult. The extreme opinion divide affected not only society as a whole but also destabilized multiple families and local communities.

In this paper, we present a novel approach, in which we combine rich social media data with the power of methods of statistical physics, to study political opinion polarization mechanisms, seeking to understand what mechanisms turned Ukraine into a irreconcilably polarized state. Though the integration of the two approaches has been increasingly discussed \cite{HelbingBalietti11,BirkinMalleson12}, there are only few studies so far that have actually attempted to combine these \cite{Serranoetal15,Randetal15,goncalvez.b:2011,weng.l:2012,takayasu.m:2015}. None of them, however, studies polarization and integrates the unstructured data analysis comprehensively with the computational model. In the past studies there has been a methodological gap between the social media analysis and agent-based modeling, which has limited the relation between the two to rather qualitative statements. We are presenting an approach here that elaborately links the two approaches. Combining them has the advantage to parameterize the computational model and validate it by empirical evidence and on the other hand to make use of the rich social media data in a theory-guided way beyond mere descriptives. This holds the potential to gain new insights into the underlying social mechanisms of polarization. In the following, we will describe a new computational model of polarization, a 2D lattice agent-based model, that is informed by the theoretical work on polarization discussed below and that brings in the spatial dimension and emphasizes the role of regional differences. We then look at the empirical polarization dynamics in the Ukrainian Twittersphere, before parameterizing the model with the analyzed Twitter data, hereby validating the model. The combination of the two approaches reveals the important role that emotional intensity levels play in polarization thus far not sufficiently accounted for by classical theoretical or empirical studies of polarization. The few studies that investigate the role of emotions in polarization have usually focussed on very specific emotions, e.g. ``self-conscious" emotions like pride and embarrassment/shame and showed that these emotions can reinforce conformity and polarization \cite{Suhay15}. In this paper, we do not focus on specific emotions, but, rather examine the role of emotional intensity levels and show how these emotional intensity levels can be a decisive driver in polarization.   

\section{Opinion Polarization}
Opinion polarization has been intensely studied over the last few decades, initiated by the observation that groups tend to adopt positions that are more extreme than the initial individual positions of its members \cite{Isenberg86,MoscoviciZavalloni69}. One explanation of this phenomenon is based on the Social Comparison Theory, which suggests that people want to be perceived in a more favorable way than what we perceive to be the average tendency. Through observations they determine what the average tendency is and then they express a slightly more extreme opinion than the perceived average opinion \cite{Isenberg86,DeGroot74,Sunstein02}. There is clear evidence for this assumption from numerous experimental studies \cite{Isenberg86,BaronRoper76,Myersetak77}. Moreover, due to the homophily phenomenon \cite{McPhersonetal01}, which states that people are more likely to interact with those who are similar to them with respect to socio-economic background \cite{Currarinietal09} as well as attitudes \cite{Dandekaretal13}, people are more likely to socially compare themselves to similar peers. Group polarization phenomena are also explained drawing on the Persuasive Argument Theory, which states that people are more likely to change their opinion when presented with persuasive arguments. Group polarization can occur when the group discourse is manipulated through biased information or misinformation, exposing a group to false persuasive arguments \cite{Isenberg86,Sunstein02,Burnstein82}. The theory is strongly supported by empirical evidence from numerous experimental studies \cite{Isenberg86,EbbsenBowers74,VinokurBurnstein78}. The two mechanisms, social comparison and persuasive argument usually co-occur. For instance, persuasive arguments in an environment of biased information can further push an individual to adopt a more extreme attitude than expressed in the group they compare themselves to. Another mechanism with respect to polarization dynamics is the Biased Assimilation \cite{LordLepper79} in opinion formation processes, which maintains that people are likely to keep their original position and draw support for it if confronted with mixed or inconclusive arguments. This mechanism, which again is empirically well established \cite{Milleretal93,Munro02,TaberLodge06}, shows that people are only to a limited extent open to change their opinions and this can contribute to polarization dynamics. In fact, Dandekar et al. \cite{Dandekaretal13} show that the two earlier described mechanisms and in particular the social comparison mechanism, is not sufficient to produce polarization, the biased assimilation mechanism has to be added.

Polarization processes have been empirically investigated through experimental studies as mentioned previously and to a lesser extent through survey based studies \cite{Baker05,Dimocketal14}. Furthermore, applied statistical physics and agent-based modeling approaches \cite{Dandekaretal13,BaldassarriBearman07,Castellanoetal09,Masetal13} have been used extensively to study polarization mechanisms through computer simulations. Prominent are for instance bounded confidence models of opinion dynamics, stochastic models for the evolution of continuous-valued opinions within a finite group of individuals that explore conditions for consensus and opinion fragmentation, introduced by Deffuant et al. \cite{Deffuantetal00} and further elaborated by numerous studies \cite{HegselmannKrause,Fortunatoetal05,Lorenz07}. We will draw inspiration from these models as well. The challenges of using statistical physics models for modeling social phenomena are known. The tractable models are usually oversimplified and too general and thus lack flexibility required to reflect the features of a particular social phenomena. Moreover, while computational studies are rigorous in investigating the specific mechanisms and dynamics of polarization they often lack empirical foundation and thus it remains often unclear to what extent these often abstract models accurately represent phenomena we see in real world. More recently, social media (e.g. Twitter and Facebook) data have been increasingly used to study opinion polarization \cite{Bakshyetal15,Conoveretal11,YardiBoyd10,AsherBandeiraSpaiser}. Gruzd and Tsyganova \cite{GruzdTsyganova14}, for instance, use Ukrainian Vkontakte data to demonstrate the split between the two political camps in Ukraine (pro-East vs. pro-West) during the Maidan protests. Twitter has also played a quite important role as a contested public debate arena throughout the crisis in Ukraine \cite{Stern14,Ronzhyn14}. Studies with social media data have revealed the strong effect of homophily in online social networks, which may lead to phenomena like the echo chamber \cite{Colleonietal14,Barbera15,AsherBandeiraSpaiser}, where opinions are amplified through communication and repetition inside an ``enclosed'' social system. Echo chambers can prevent people from noticing contrary persuasive arguments and they skew the perceived average that people take into consideration in social comparison processes. Though social media data is potentially rich, i.e. fine-grained, time-resolved, relational, geo-coded, etc., without an explicit theoretical underpinning, the studies of polarization on social media remain usually rather descriptive. By combining rich social media data with the power of methods of statistical physics a better understanding of specific political opinion polarization mechanisms is sought in this paper.

\section{Methods}

\begin{figure*}
\begin{center}
\includegraphics[scale=0.45]{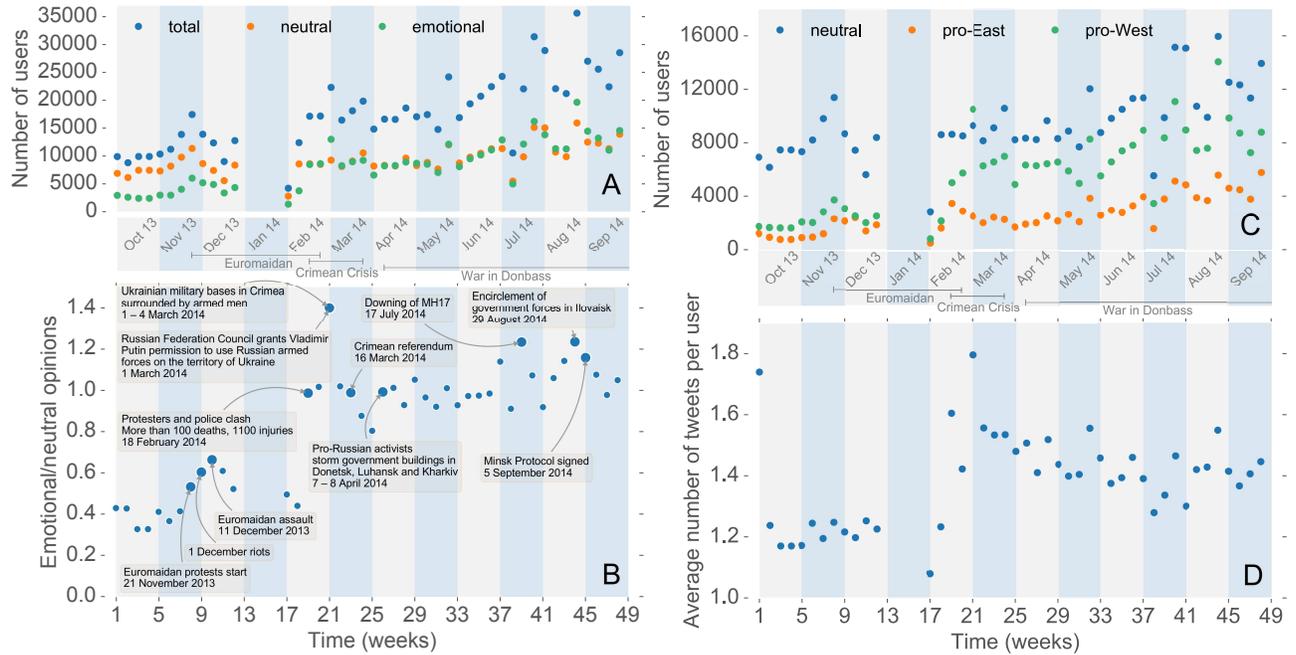}
\end{center}
\caption{(A) Time series plot of number of tweets, in total, neutral and emotional. (B) Time series plot of emotional/neutral tweets ratio with key political events tags. (C) Time series plot of number of average (weekly) users' opinions, in total, pro-West and pro-East. (D) Average weekly number of tweets per user. No data was available for January 2014 (see Materials and Methods and the electronic supplementary material) therefore the data gaps in plots A-D between the weeks 13 and 17.}
\label{twitter1}
\end{figure*}

\subsection{Political opinion mining of Twitter data}
We used the archived Twitter API Streaming data from October 2013 to September 2014 provided by the Archive Team \cite{archive}. The archived data is the freely available Twitter Streaming API Spritzer Sample, which collects 1\% of all public tweets in real-time. The Twitter Streaming API Spritzer Sample allows for unfiltered data analysis and hence for capturing the full discourse picture comparing to a more narrow, focussed (e.g. based on certain hashtags, user networks etc.) approach facilitated by data using the Twitter REST API \cite{Leeetal17} that would disregard discourse contributions beyond the specified search queries. The data for January 2014 was missing and we did not find any other archive providing this data. The tweet data is stored in JSON format. The data was processed and analyzed in Python. Specifically, the data was filtered for Russian/Ukrainian language (excluding Twitter users who specified being from Russia), cleared of SPAM, and then filtered for political content with an extensive set of keywords (see the electronic supplementary material for details). We used the tweets to determine the political affiliation of each Twitter user using a set of keywords indicating neutral, pro- or anti-West attitudes and a set of keywords indicating neutral, pro- or anti-Russian attitudes (see the electronic supplementary material for details). Moreover, we conducted a sentiment analysis of the tweets, in order to determine the political affiliation for previously seemingly neutral Twitter users. For this purpose we utilized the SentiStrength system for automatic sentiment analysis, built an Ukrainian sentiment words dictionary (see the electronic supplementary material) and extended the SentiStrength Russian sentiment words dictionary to make it equivalent to the Ukrainian sentiment words dictionary (see the electronic supplementary material for details). Our classification algorithm scanned all tweets for each user and checked whether the tweet contained any of the keywords and any of the sentiment words specified and calculated overall scores for political ``West'' and ``East'' affiliation as well as sentiment scores (see the electronic supplementary material for details). For the geo-plots in Fig. \ref{twitter2}C,D, we focused on the data of the last week in September 2014. We scanned the Twitter data of each user for two possible geographical information, the value of their ``location'' tag and/or geo/place ``coordinates'' tag which is a latitude and a longitude coordinate value. The vast majority of users have the default option ``geo\_enabled:false'', thus do not provide precise geographical information attached to the tweet, a few more provide profile ``location'' information, but overall, geographical data is often not specified and thus missing. We therefore ended up with only 493 Twitter users for whom we had a geographical information. The ``West'' and ``East'' political affiliation scores of these 493 Twitter users were plotted in Fig. \ref{twitter2}C,D (see the electronic supplementary material for details).

\subsection{Bounded Confidence XY Model of Opinion Polarization}
We model opinion dynamics in an assembly of interactive agents placed on a two-dimensional lattice of finite size, which emulates a regional opinion distribution and localization. Each opinion is represented by a vector that can freely rotate in plane, similar to that in the $XY$ model of a magnet, and is characterized by the length $p$ (emotional intensity, i.e. fervent strength of an opinion) and polar angle $\theta$ (orientation). This allows us to model a continuous opinion spectrum with respect to the specified direction by a cosine of the angle between the vectors, which can vary in the range between -1 (agent opposes the opinion) and 1 (agent fully supports the opinion). Each agent can interact with a fixed number of nearest neighbors. The social interactions are introduced via local mean field, similar to the Vicsek model \cite{vicsek.t:1995}: the new orientation of each vector is calculated as a mean of the average direction of the neighbors and agent's own orientation:
\begin{equation}\label{eq:interaction}
    \theta_i (t+\Delta t) = \tan^{-1}\left [ \frac{1}{N} \frac{ \sum_{j, |j-i|\leq r}{p_j \sin(\theta_j (t))}} {\sum_{j, |j-i|\leq r}{p_j \cos(\theta_j (t))}}  \right ]+\xi(\eta,t),
\end{equation}
where $N$ is the number of interacting agents, $\xi$ is the angular noise variable uniformly distributed in the interval $[-\eta/2, \eta/2]$ and $\eta$ is the noise strength. The noise is added to model the level of conformity of the individual: zero noise, $\eta=0$ corresponds to full conformity, $\eta = 2 \pi$ allows an individual to deviate from the group opinion by an unlimited value. A contribution of each interacting agent is weighted by the emotional intensity of its opinion $p$. The interaction has a limited range $r$, which sets the number of nearest neighbors considered. In 2D, interaction range $r=2$ corresponds to 24 interaction peers, $r=3$  gives 48 peers, etc (see the electronic supplementary material for details). In addition to this, the interaction is selective so that the vectors align only with those neighbors, whose orientation (opinion) deviates by an angle less than some fixed value $\alpha$ from their own opinion vector. This rule is inspired by the bounded confidence model, or Deffuant model  \cite{Deffuantetal00}, and allows one to imitate systems with different levels of opinion tolerance \cite{romensky.m:2014}. Scaled value $\alpha/\pi$ denotes a fraction of the opinion spectrum that is taken into account by each individual, e.g. $\alpha /\pi= 0.1$ corresponds to 10\% closest opinions taken into account. Note that our model does not explicitly include any intrinsic preference to agent's own opinion (i.e. orientation of the opinion vector $i$ in the previous time step) because using a relatively small values of $\eta$ and $\alpha$ already allows to achieve a socially realistic behavior.

We model a finite system with the boundary conditions set by two fixed rows of agents on each side. The boundary vectors are fixed according to the following scheme: the left and upper rows are oriented to the left ($\theta=\pi$), imitating a bias towards ``West'' and the right and bottom rows are fixed to point in the opposite direction ($\theta=0$), thus mimicking a bias towards ``East''. When other agents interact with the boundary vectors this introduces a spatially dependent local bias that can account for geographical inhomogeneity in opinions, thus imitating cultural or ethnic differences, information bias, etc. (see the electronic supplementary material for details; see also the interactive model \cite{applet}).

These model settings reflect the theoretical assumptions about polarization described above. The orientation represents political attitudes and the local mean field calculation of new orientations for each agent represents the social comparison theory assumptions. The homophily effect is included through the bounded confidence feature, where agents align only with those neighbors, whose political orientation is similar to their own. Furthermore, the biased assimilation mechanism is reflected in the noise variable that determines the agents' willingness to change opinion (conformity level). Finally, the persuasive argument theory assumption, in particular with respect to the contribution of biased information to polarization, is simulated via the boundary conditions. We have added a further parameter to our model, the emotional intensity, i.e. vector length, that represents the level of emotional strength and vehemence of a political opinion. This parameter reflects results we have obtained from Twitter data sentiment analysis discussed in the next sections. It also builds on recent computational models and big data analysis of opinion dynamics \cite{Sobkowicz12,Sobkowicz13}, which showcase the importance of emotions, and in particular of negative emotions, for people's engagement in political debates and for opinion formations and changes.

We modeled evolution of the opinion spectrum in the described system starting from randomized initial distributions of agents' emotional intensity and orientations based on general uniform distributions. The emotional intensity levels for each agent were kept fixed in each simulation while the orientations evolved due to noise and interactions. For each opinion spectrum extracted from Twitter data, we performed a simulation until a steady state was reached. After that, the following statistics were collected: steady state distribution of the opinion along the ``East-West'' scale, as defined by the boundary conditions, mean order parameter, and bimodality index.

We should note here that in this setup the week-by-week series of calculated properties do not reflect the real time, nor the actual system's dynamics as the history of the individuals as well as previous steady states are ignored. To follow the variation of collective properties, ideally one should look at the evolution of each user's opinion and derive the group behavior from the corresponding statistics. For this purpose, one would need to either parameterize the individual opinions from the empirical data directly or solve the inverse problem and introduce the variation of opinions based on the instantaneous statistical averages. As we could reach only a random sample of the tweets, it was not possible to follow the former route and track individual users. The empirical data we have are discontinuous. Moreover, we have no appropriate model for individual psychology. Therefore, we decided not to introduce an artificial evolution of the opinions. As the individual history is lost, we can only follow the variation of the averages corresponding to the snapshot of Twitter data. 

\subsection{Model parameterization with Twitter data}
To parameterize the Bounded confidence XY model we used weekly distributions of the overall users' emotional intensity scores. For each week, the overall emotional intensity of each user was defined from the mean of the average sentiment scores on a continuous scale from 0 to 5, with 0 corresponding to a neutral average user's opinion and 5 reflecting an extremely emotional average opinion. This measure did not contain any information about political affiliation, so that each non-zero value could correspond to either pro-West or pro-East user's political attitude. The data was sampled with a bin size of 1 for all non-zero opinions and a resulting discrete distribution was normalized by the total number of users per week giving probabilities for each discrete value of overall emotional intensity (0 to 5). This distribution was then applied to the simulated system to define the length $p$ of each opinion vector. In each simulation, values of $p$ were assigned to agents at random, according to obtained discrete weekly distributions of overall emotional intensity, and kept constant throughout the simulation. The simulated system consisted of 14641 agents placed in nodes of $120\times120$ lattice. We performed at least $10^6$ update cycles to determine the structure of the steady state. Each statistics was averaged over at least 5 independent runs. We computed the bimodality coefficient as $\beta={(\gamma^2+1)}/{k}$, where $\gamma$ is the skewness and $k$ is the kurtosis of weekly distributions of average opinion scores in Twitter data or weekly distributions of cosines of orientations $\theta$ of opinion vectors in simulations. 

\section{Results}

\begin{figure*}
\begin{center}
\includegraphics[scale=0.5]{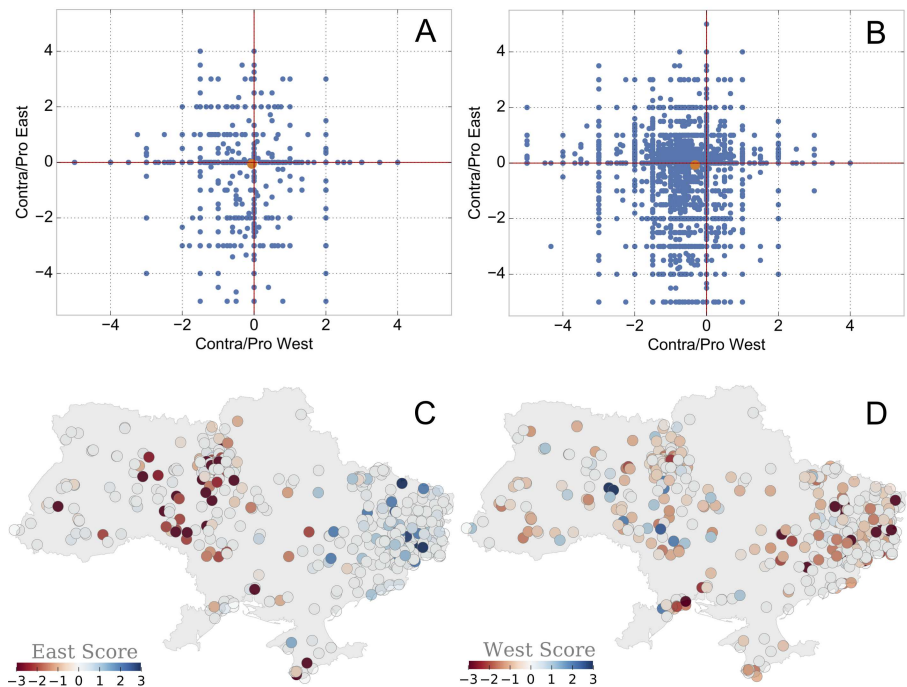}
\end{center}
\caption{(A) Pro-against-European (West)/pro-against-Russian (East) opinion space plot for the first week of October 2013. (B) Pro-against-European (West)/pro-against-Russian (East) opinion space plot for the second week of August 2014. The points in plots C and D are Twitter users based on their two scores, the orange dot is the average. (C) Twitter users (dots) colored according to their ``East'' affiliation score and plotted over the Ukraine map using Twitter geo, place or location information. (D) Twitter users (dots) colored according to their ``West'' affiliation score and plotted over the Ukraine map using Twitter geo, place or location information.}
\label{twitter2}
\end{figure*}

\subsection{Opinion Polarization in the Ukrainian Twittersphere}
To validate the model assumptions, we used Twitter Streaming data from October 2013 to September 2014 provided by the Archive Team. We use Twitter data because it provides fine-grained, time-series and rich data on recent political opinion dynamics in Ukraine, otherwise not available. To determine the political affiliation of the Twitter users in our data, we used a political affiliation classification procedure based on keyword and sentiment analysis suggested and tested by Spaiser et al. \cite{Spaiseretal} (see the electronic supplementary material for details). As a result, every Twitter user was assigned two scores, a ``West'' score and an ``East'' score, representing their political position in a pro-against-European (West)/pro-against-Russian (East) opinion space (see Fig. \ref{twitter2}A,B) and an emotional intensity score, combining their sentiment analysis scores.

Our Twitter analyzes show that political polarization did indeed take place in the Ukrainian Twittersphere between October 2013 and September 2014. Fig. \ref{twitter1}A shows a discontinuous increase in emotional users in February 2014, while the total number of users and the number of neutral ones increased steadily. Fig. \ref{twitter1}B depicts a ratio of emotional/neutral opinions, where a jump from values of about 1.0 to about 1.4 is visible around February as well. We added to this figure the key political incidents in Ukraine during this year, so the discontinuous changes in the data can be related to actual political events. This shows that the biggest discontinuous change in February took place around the time when the Maidan protests escalated and 100 people died on a single day. These results inspired the inclusion of emotional intensity levels as a parameter in the computational model. In addition to an abrupt increase in the total number of users (Figs. \ref{twitter1}A,C), users' involvement in the topic also changed discontinuously (Fig. \ref{twitter1}D). The dynamics of the average number of tweets per user completely resembles that of the emotional level of tweets (Fig. \ref{twitter1}B), with a characteristic jump around week 19 (February 2014).

\begin{figure*}
\begin{center}
\includegraphics[scale=0.45]{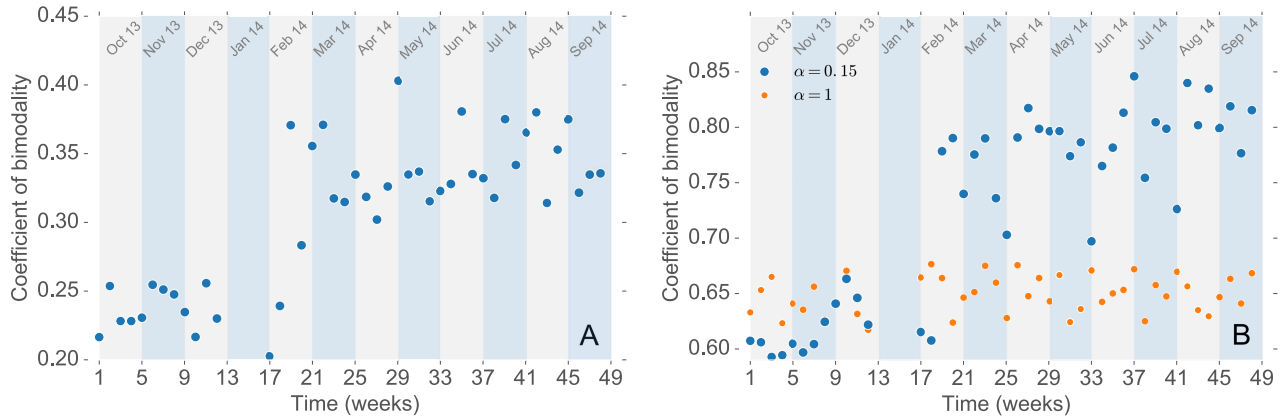}
\end{center}
\caption{(A)Time series plot of coefficient of bimodality of Twitter users' opinion distribution. (B) Time series plot of coefficient of bimodality of vector orientations in simulations ($\eta=0.12$).}
\label{twitter_vs_sim}
\end{figure*}

Fig. \ref{twitter2}A,B moreover shows the polarization in a continuous two-dimensional space defined by pro-against-European (West)/pro-against-Russian (East) scores. This figure shows that the polarization resulted in appearance of two main opinion clusters, those who are pro-East and against-the-West on the one side and those who are against East but who are also quite critical of the West on the other side (see the electronic supplementary material for details). This in fact reflects most current surveys that show the disappointment of many Western oriented Ukrainians with the pro-West Poroshenko government \cite{Gallup15}.

Our Twitter data analysis confirms that political affiliation follows the expected geographical pattern. People who are pro-East and against-West are more likely to be located in the Eastern and Southern parts of Ukraine, while Ukrainians with rather an against-East and (critical) pro-West attitude are to be found in the Western and Northern territories (see Fig. \ref{twitter2}C,D). This analysis is a validation of our classification procedure, since the geographical distribution of the Twitter users with their respective political affiliation scores matches the actual geographical political camp distributions in Ukraine (see Fig. \ref{bbc}).

\subsection{Validating the Bounded Confidence XY Model of Opinion Polarization with Twitter data}
To quantify the opinion divide, we present here the bimodality index for the opinion spectrum (see Materials and Methods for the definition), as we found it to be most sensitive to the changes of the spectrum, therefore, the best integral characteristic of the observed behavior. The changes in the bimodality coefficient over 12 months from October 2013 are shown in Fig. \ref{twitter_vs_sim}A,B. In the Twitter data the bimodality keeps as low as 0.2 to 0.26 from the start of observation until week 18 (February 2014) but then demonstrates sudden increase to 0.35 to 0.40 within the next two or three weeks, after which it stays high, and the original value is never restored. This sudden increase of the bimodality reflects a formation of two distinct political camps and clearly shows that the opinion divide has suffered a significant deepening in this short period. Moreover, we see that the deepening was non-recoverable in the short term. The computational model, parameterized with the Twitter emotional intensity spectrum for each respective week, captures this behavior well and shows a similar qualitative trend. In a simulation with $\alpha = 0.15 \pi$, the bimodality coefficient jumps from ca. 0.60 -- 0.65 to 0.8 -- 0.85 during the same period. The quantitative difference in the values is mostly due to the use of cosine function. We should note that, although the model is parameterized by the Twitter data, it is not bound to reproduce the distribution, as the vectors are allowed to change the orientation, and this behavior follows only from the specific anisotropic interactions between the agents. We repeated the simulations with a different interaction parameter $\alpha=1.0$, corresponding to agents without any resistance to opinion change and found no jump in the bimodality. Therefore, the opinion divide is conditioned by both the restriction of confidence and by conformity (noise) levels, thus confirming the homophily and biased assimilation assumptions. Moreover, the drastic change in bimodality corresponds to the sudden increase of the emotional intensity, which we noted in the data in Fig. \ref{twitter1}B.

\begin{figure*}
\begin{center}
\includegraphics[scale=0.37]{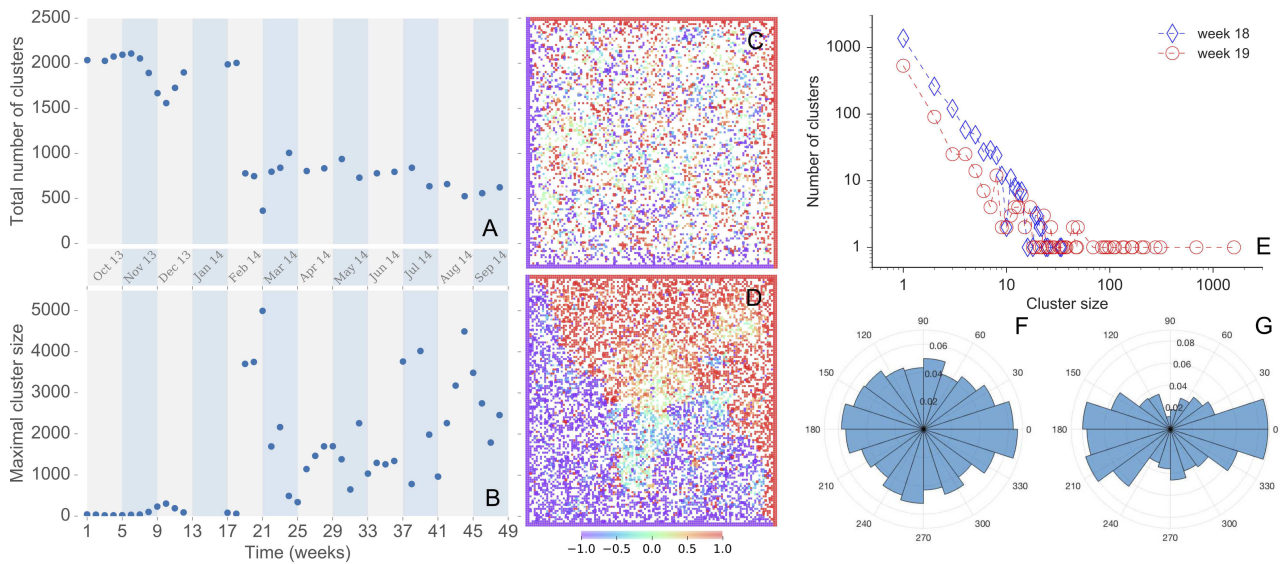}
\end{center}
\caption{(A)Time series plot of average number of clusters in simulations. (B) Time series plot of maximal cluster size in simulations. (C) Simulation snapshot for week 18, February 2014. (D) Simulation snapshot for week 19, February 2014. Each square in plots C and D represents an individual agent; size of a square is proportional to emotional intensity; color of each square denotes cosine of an orientation angle for each spin, with -1 and 1 corresponding to orientation towards ``West'' and ``East'', respectively. (E) Log-log plot of distribution of cluster sizes in simulations for week 18, February 2014 (blue) and week 19, February 2014 (red). (F) Simulation cluster orientation diagram for week 18, February 2014. (G) Simulation cluster orientation diagram for week 19, February 2014. Orientation of clusters in plots F and G is shown in degrees; 180 and 0 degrees denote orientation towards ``West'' and ``East'', respectively; length of each bin reflects cluster probability. Simulation parameters in plots A--G are $\alpha=0.15$ and $\eta=0.12$.}
\label{twitter_vs_sim2}
\end{figure*}

\subsection{Onset of Opinion Clustering and Formation of Territorial Domains}
The increase in emotional intensity leads to important consequences in the spatial dimension. In the weeks before and early at the outbreak of the crisis, the simulations display large diversity of the opinions (Fig. \ref{twitter_vs_sim2}C) characterized by the large number (Fig. \ref{twitter_vs_sim2}A) of small (Fig. \ref{twitter_vs_sim2}B) opinion domains (clusters). This picture, however, changes further into the crisis, around week 19, when a smaller number of larger opinion domains is formed. While clusters rarely exceed 100 agents before the critical weeks 17-18, starting from week 19 we observe clusters of up to 5000 agents and the numbers rarely drop below 1000. That the behavior becomes expressly collective is consistent with the rise of emotional intensity and of the fraction of involved agents (those with non-zero valence), which increases the number of interactions within each individual's circle and thus the local aligning field. This is further confirmed by Fig. \ref{twitter_vs_sim2}F,G showing the polar opinion histogram for the two critical time points, week 18 and 19 obtained from simulation analyses. The distributions are visibly gravitating towards 0 or 180 degrees (``East'' and ``West'') in both sets. The fraction of neutral opinions drops from week 18 to week 19 and the distribution of opinions in each subgroup becomes very narrow.

Simulation analysis moreover shows that the opinion divide induces also territorial splitting. Subfigures Fig. \ref{twitter_vs_sim2}C,D show the in-plane opinion distribution. We plotted the opinions as predicted by the model just before the jump in the bimodality (week 18 -- C) and immediately after that (week 19 -- D). The change in the distribution between these points is dramatic: while the predicted data for week 18 show a merely uniform distribution of both ``East'' and ``West'' orientations, the picture for week 19 features two distinct clusters with predominant ``West'' orientation in the lower left corner and domination of the ``East'' orientation in the top right corner. These orientations correspond to vector directions in the preset boundary conditions. Each domain contains practically no opposing opinion, as they are squeezed out to the periphery and then to the opposing domain as the steady state develops. We can see small islands of mixed/neutral opinion in the middle of the simulation domain. The prediction of the territorial divide matches also well the geo-location data shown in Fig. \ref{twitter2}C,D. We should stress however that the biased boundary conditions alone are not sufficient to produce any large domains even in the system with limited confidence (Fig. \ref{twitter_vs_sim2}C, see the electronic supplementary material for details), although they definitely facilitate this collective behaviour.

The simulation allows us to analyze the steady states of the system and provides insights into the mechanisms of sudden onset of polarization, clustering, and territorial splitting (see the electronic supplementary material for details). We in particular examined varying levels of noise, restriction angles and vector lengths. High noise (low conformity) corresponds to a globally disordered behaviour (without any prominent consensus or polarization) and the range of higher $\alpha$ allows only states with polar order (global consensus). Smaller restriction angles, $\alpha < 0.4 \pi$, on the other hand produce regions of prevalence of the bipolar states, thus structure the system in a polar or bipolar way. A combination of small restriction angle (strongly bounded confidence) and high noise (low conformity) does not produce any global order and is rather unrealistic since at these conditions the system represents a set of selective but randomly vacillating agents. Polarized states are generally only possible at low noise (high conformity) and small restriction angles (strongly bounded confidence). The changing level of emotional intensity pushes the boundary between the polarized and non-polarized societies moreover outwards, thus extending the range of the polarized states, and brings the originally weakly polarized society to a highly polarized one.
This splitting resembles in appearance a phase separation in dissimilar liquids (e.g. oil in water). The important difference of our system from the liquid state systems is that the agents are not dissimilar from the beginning but the dissimilarity and effective repulsion between the opposite arises from strong social interactions that dictate cohesion between the like opinions. Another important observation here is that the transition is driven primarily by the increase of emotional level while the other parameters (conformity, confidence) stay constant. We should stress that the source of and the original direction of the emotional agents were not crucial. The key properties of the model that determine the nature of the steady states (non-polar, polar or bipolar) are the high conformity and strongly bounded confidence.

\section{Conclusion}

We have analyzed the opinion dynamics over a recent period of political unrest in Ukraine. Based on the Twitter data, we registered an onset of emotional intensity of tweets that corresponded to rising levels of involvement of the population in the political feud, fueled by the action of the government, collisions of the opposing groups, and foreign military activities. The escalating opinion divide around the time became apparent among others in the jump of the opinion bimodality index. We proposed an agent-based lattice model to study political polarization as a collective behavior including the spatial dimension of polarization. By parameterizing the model with Twitter data at distinct time points, we predicted the onset of collective behavior and territorial splitting of the opinion. We demonstrated that the tendency of territorial splitting is conditioned by the high conformity and homophily in the society and is driven by the growth in emotional intensity. Our analyzes demonstrate clearly the importance of emotional intensity for polarization, a factor that has been largely ignored thus far in classic theoretical and empirical literature on polarization with a few noteworthy exceptions as discussed earlier. Specifically, while our analyzes confirm the importance of social comparison, homophily, persuasive argument and biased assimilation mechanisms and their specific interactions for polarization, they also show that these mechanisms are not sufficient to ignite the extreme societal polarization we can observe for instance in Ukraine. The emotional intensity is a key reinforcement mechanism that has to be added.

Our analysis seem also to suggest a link between polarization and separatist trends. The polarization dynamics that we establish for February 2014 in the data and simulation increases in fact further around April and May 2014, when the self-declared Donetsk and Luhansk People's Republics were formed in the South-East of Ukraine. This seems to suggest that the opinion split may facilitate the separatist trends on its own. The observed phenomenon is not unique to Ukraine, and similar processes of deepening polarization leading to separatism are well known elsewhere (e.g. Northern Ireland). In most cases, however, the separatism is related to more obvious ethnic, racial, or religious differences between the communities, while in Ukraine the division is more subtle and roots in small cultural differences, which were artificially enhanced by external factors. And our results show how dangerous targeted agitation can be when it is backed by modern information warfare techniques \cite{Cottiero15,Aro2016} as it boils emotions making polarized world views increasingly irreconcilable.

\subsubsection*{Ethics}{Tweets were collected and analysed in accordance with the Twitter Privacy Policy and the Twitter Terms of Service.}

\subsubsection*{Data accessibility}{The datasets supporting the conclusions of this article are provided by the Archive Team \cite{archive}.}

\subsubsection*{Authors' contributions}{MR: Development and Testing of the Bounded Confidence XY Model, Parametrization and Testing of the Model with Twitter Data, Development of Ukrainian dictionary for sentiment analysis, Preparing graphics, Writing the paper, VS: Processing and Analyzing Twitter data, Preparing graphics, Writing the paper, TI: Development of the Bounded Confidence XY Model, Writing the paper, VL: Development of the Bounded Confidence XY Model, Writing the paper.}

\subsubsection*{Acknowledgements}{The authors thank Ian Harper for help with programming and testing of the web applet.}

\vfil

\widetext
\clearpage
\begin{center}
\textbf{\large Supplementary Material for: Polarized Ukraine 2014: Opinion and Territorial Split Demonstrated with the Bounded Confidence XY Model, Parameterized by Twitter Data}
\end{center}
\begin{center}
\text{Maksym Romenskyy,\textsuperscript{1,2} Viktoria Spaiser,\textsuperscript{3} Thomas Ihle,\textsuperscript{4} Vladimir Lobaskin\textsuperscript{5}}
\end{center}
\begin{center}
\textit{\textsuperscript{1}Department of Life Sciences, Imperial College London, London SW7 2AZ, UK}
\textit{\textsuperscript{2}Department of Mathematics, Uppsala University, Box 480, Uppsala 75106, Sweden}
\textit{\textsuperscript{3}School of Politics and International Studies, University of Leeds, Leeds LS2 9JT, UK}
\textit{\textsuperscript{4}Institute of Physics, University of Greifswald, Felix-Hausdorff-Str. 6, Greifswald 17489, Germany}
\textit{\textsuperscript{5}School of Physics, University College Dublin, Belfield, Dublin 4, Ireland}
\end{center}
\begin{center}
\text{(Dated: \today)}
\end{center}
\setcounter{equation}{0}
\setcounter{figure}{0}
\setcounter{table}{0}
\setcounter{page}{1}
\makeatletter
\renewcommand{\theequation}{S\arabic{equation}}
\renewcommand{\thefigure}{S\arabic{figure}}
\renewcommand{\bibnumfmt}[1]{[S#1]}
\renewcommand{\citenumfont}[1]{S#1}

\section*{Supplementary Methods}
\subsection*{ Further Model Specifications}

\begin{figure}[!h]
\begin{center}
\includegraphics[scale=0.6]{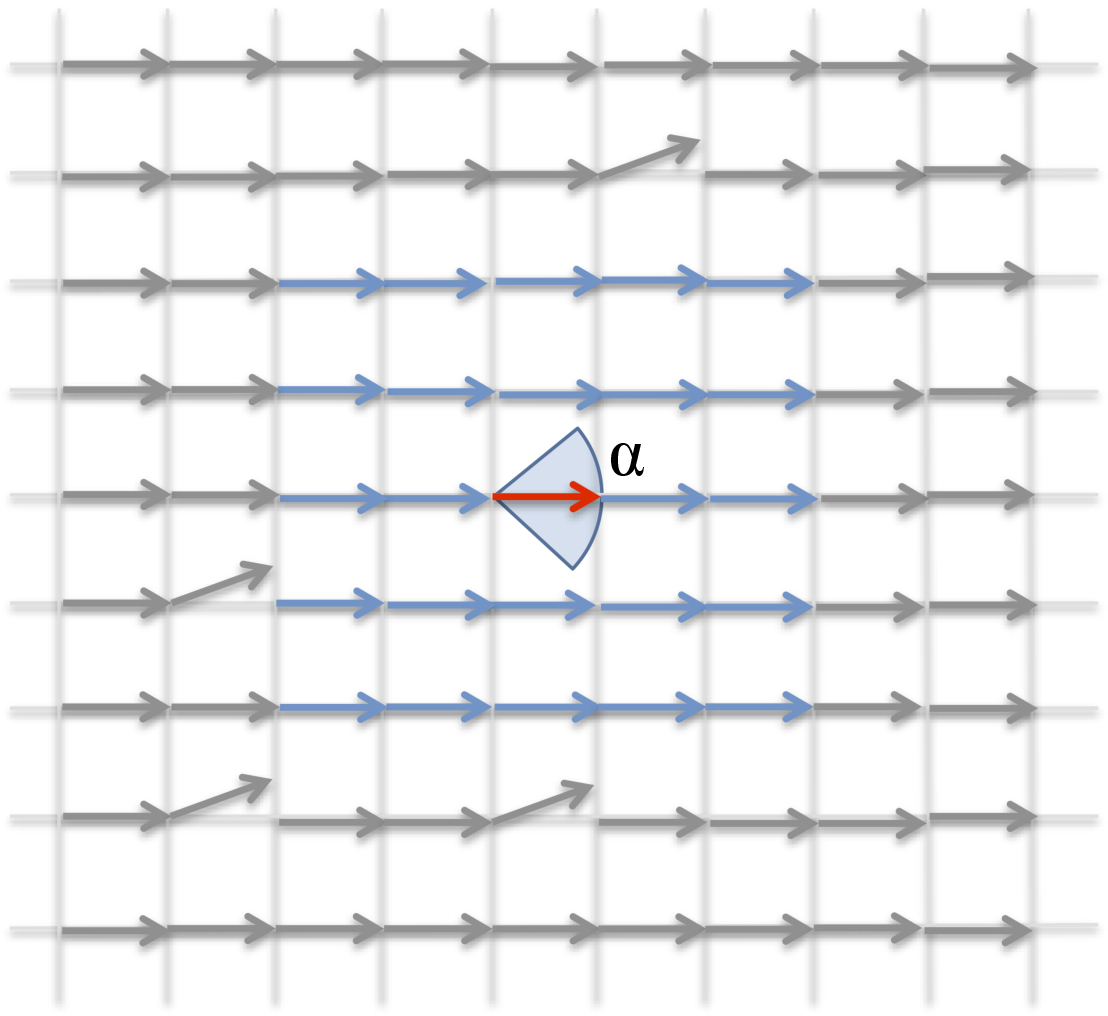}
\end{center}
\caption{The interaction parameters in the Bounded Confidence XY model with $r=2$. The focal opinion vector (red arrow) interacts only with those neighbors whose relative orientation is less or equal to $\alpha$ and who are located within the two nearest rows/columns (blue arrows).}
\label{interaction}
\end{figure}

Supplementary Figure \ref{interaction} illustrates interactions in the Bounded Confidence XY model. The red opinion vector interacts with its 24 neighbors located within two nearest lattice nodes in any direction $r=2$. In this illustration, each of the 24 neighbors has orientation relative to the focal (red) vector less than $\alpha$ and therefore all 24 agents contribute to future orientation of the focal vector. For simplicity, all vectors in Supplementary Fig. \ref{interaction} are shown to have same length. In all simulations, the initial orientations of all agents except the ones at the system's boundary were drawn randomly from the uniform distribution in the interval $[-\pi, \pi]$.

To quantify formation of territorially isolated domains in our model, we performed a simple cluster analysis. We define a cluster based on two criteria: distance and relative orientation between the agents.
Therefore, a cluster is a set of connected agents, each of which is within the cutoff distance (defined by $r$) from one or more other opinion vectors from the same cluster and a relative orientation between any two neighbors in the cluster is less or equal to restriction angle $\alpha$. Conversely, two agents will not belong to the same cluster, if there is no continuous path on the neighbor network leading from the first agent to the second or if this path is broken because the angle between two neighbors is larger than $\alpha$ and hence the opinion vectors do not interact. For each update step we calculated size of each cluster, maximal cluster size and total number of clusters.

\begin{figure*}
\begin{center}
\includegraphics[scale=0.5]{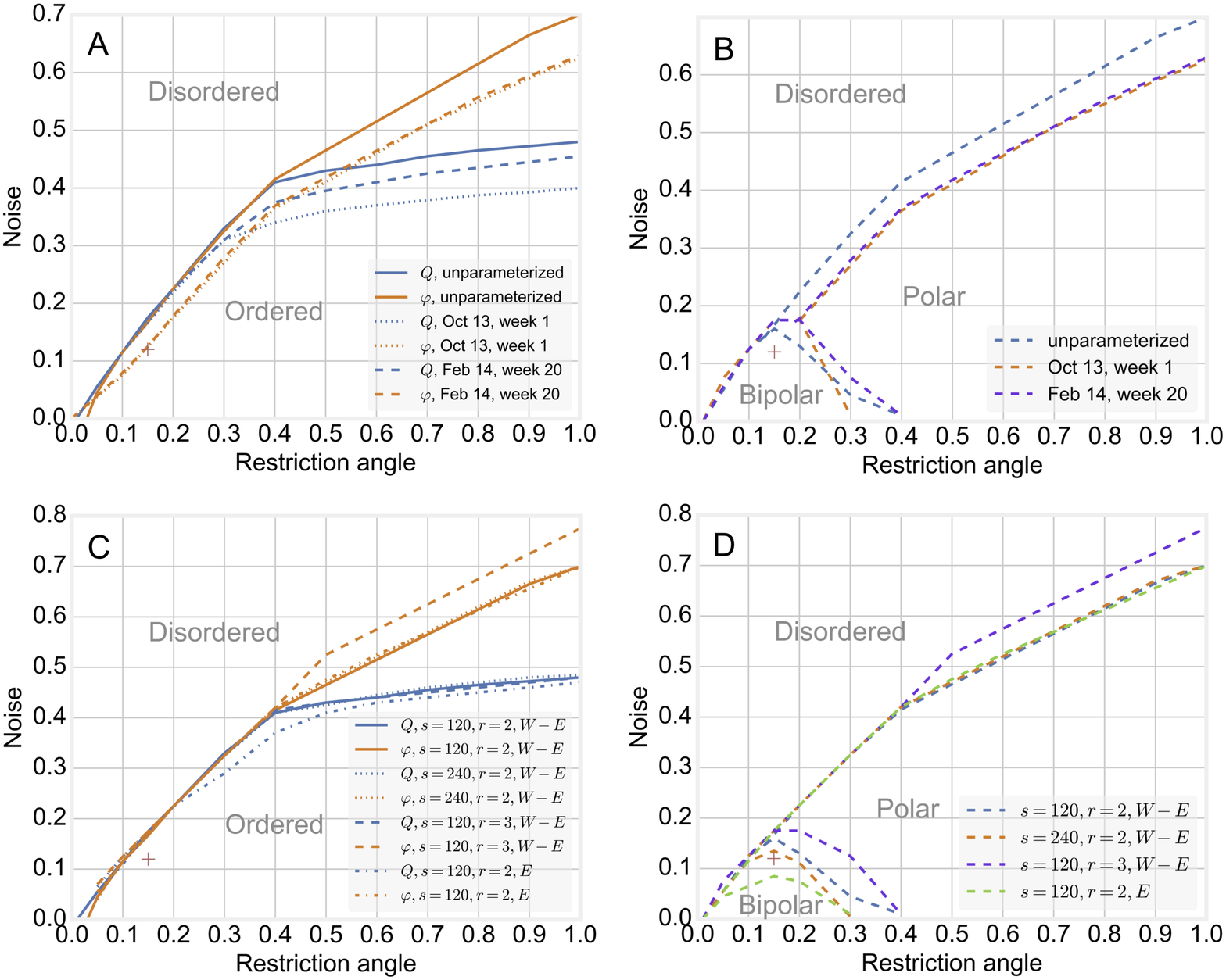}
\end{center}
\caption{Phase behavior of the Bounded Confidence XY model. (A) Phase diagrams based on polar $\varphi$ and bipolar $Q$ order parameters for the model system without and with parameterization by Twitter data. (B) Combined phase diagrams for the model system without and with parameterization by Twitter data showing regions of dominance of polar and bipolar order. (C) Phase diagrams based on polar $\varphi$ and bipolar $Q$ order parameters for non-parameterized model system at different simulation parameters (see legend). (D) Combined phase diagrams showing regions of dominance of polar and bipolar order for unparameterized model system at different simulation parameters (see legend). The red plus marker denotes simulation parameters $\alpha$ and $\eta$ for the parameterized version of the model used throughout the paper. In plots C and D, $s$ denotes size of a side of a square lattice; $E$ stands for East and means that all boundary spins are fixed to the right ($\theta=0$), $W - E$ stands for West -- East and means that the boundary spins on the left and on top are fixed to the left ($\theta=\pi$) and the boundary spins on the right and on the bottom are fixed to the right ($\theta=0$); $r$ denotes the interaction range.}
\label{phasediag}
\end{figure*}

We characterised orientational ordering in our model using two order parameters. Polar order parameter was used to quantify the average degree of opinion agreement between the agents
\begin{equation}
\varphi =  \frac{1}{N} \left |\sum_{j = 1 }^N \exp(\imath \theta_j) \right |,
\label{order_param}
\end{equation}
where $\imath$ is the imaginary unit and $\theta_j$ is the direction of each vector $j$. This order parameter turns zero in the isotropic phase, when opinions of all agents are very different from each other, and assumes finite positive values in the ordered phase, reaching unity when global consensus settles in.

To characterise opinion polarization in our model we use the following bipolar order parameter
\begin{equation}
Q=\left|\frac{1}{N}\sum_{j = 1 }^N \exp(\imath2\theta_j)\right|.
\label{apolar_param}
\end{equation}
When the two vectors are oriented perfectly collinearly, $Q=1$. Note that a perfectly polarly ordered phase is characterized by $\varphi = Q = 1$, as the polar ordering implies the bipolar ordering. A bipolarly ordered phase requires only $Q = 1$ while the polar order parameter can take any value $\varphi < 1$. Therefore, requirements for the polar order are more restrictive.

%
%

For each weekly distribution of opinions, both for the Twitter data and in simulations, we computed the bimodality coefficient (see main text, Methods) by first calculating the skewness and kurtosis of the distribution. Skewness is defined as the third standardised moment around the mean
\begin{equation}
\gamma=\frac{\mu_3}{\mu_2^{3/2}},
\label{skewness}
\end{equation}
where $\mu_2$ and $\mu_3$ are the second and the third cumulants, respectively. Kurtosis is computed as the fourth central moment
\begin{equation}
k=\frac{\mu_4}{\mu_2^2},
\label{kurtosis}
\end{equation}
where $\mu_2$ and $\mu_4$ are the second and the fourth cumulants, respectively.

\begin{figure}
\begin{center}
\includegraphics[scale=0.5]{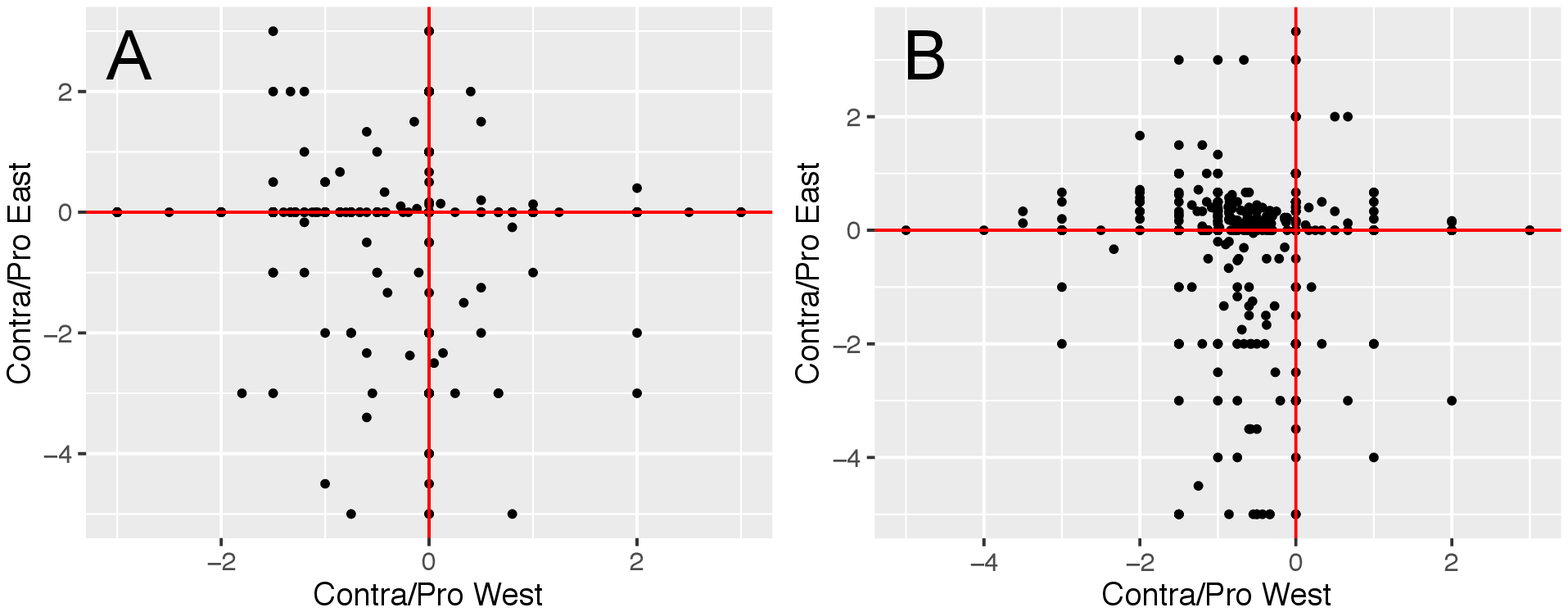}
\end{center}
\caption{(A) Panel Twitter users (black dots) in the pro-against-European (West)/pro-against-Russian (East) opinion space, data from first week in October 2013. (B) Panel Twitter users (black dots) in the pro-against-European (West)/pro-against-Russian (East) opinion space, data from first week in September 2014.}
\label{panel}
\end{figure}

The model phase diagrams, shown in Supplementary Figs. \ref{phasediag} A-D, demonstrate three different steady states of the system. The high noise (low conformity) generally corresponds to a globally disordered behavior (without any prominent consensus or polarization). The region of disordered behavior shrinks as the restriction angle $\alpha$ is increased (Supplementary Fig. \ref{phasediag} A). The lesser restriction leads to more interactions between the nearest neighbours and to the onset of order (consensus). At lower noise, below the transition line we observe either polar or bipolar structuring of the system (Supplementary Fig. \ref{phasediag} B). The range of higher $\alpha$ allows only states with global consensus, while at $\alpha < 0.4 \pi$ we see the region of prevalence of the bipolar polarized states. In other words, the polarized states are only possible at low noise (high conformity) and small restriction angles (strongly bounded confidence). The change in level of emotional intensity pushes the boundary between the polarized and non-polarized societies outwards (Supplementary Fig. \ref{phasediag} B), thus extending the range of the polarized states, and brings the originally weakly polarized society to a highly polarized one. The most important observation here is that the transition happens just due to the increase of emotional intensity while the other parameters (conformity, confidence) stay constant.
Supplementary Figures \ref{phasediag} C,D show phase diagrams for unparameterized systems at different simulation parameters. For a larger system, $s=240$ constituting of 58081 agents, the area of bipolar ordering is slightly reduced as compared to the standard system size ($s=120$) used in this study. The shrinking happens because the influence of biased boundary vectors is decreased due to larger system size. If the agents have larger interaction range, $r=3$ (i.e. each vectors interacts with up to 48 nearest neighbors), the region of bipolar ordering expands to larger restriction angles and noise values because longer correlation range becomes possible and effect of the boundary bias is more pronounced at these conditions. Finally, if all boundary vectors are fixed in one direction (in this case to the East, $\theta=0$), the area of bipolar polarization shrinks significantly but does not disappear. At these conditions bipolar ordering also displays more sensitivity with respect to noise than to restriction angle.

\subsubsection*{Twitter Data}

The archived Twitter Streaming API Spritzer Sample tweet data was stored in JSON ({\bf{J}}ava{\bf{S}}cript {\bf{O}}object {\bf{N}}otation) file format, which is the most common open standard data format to transmit data objects in asynchronous browser/server communication. The data was then processed and analyzed in Python using the Python Natural Language Text Processing Toolkit (NLTK) \citep{BirdKleinLoper09,Perkins10}. NTLK is a collection of various classes, interfaces and functions for natural language processing, including important text mining methods. It was developed at the University of Pennsylvania and is widely used in computational linguistics, machine learning, and cognitive science (see in particular \url{https://nltk.googlecode.com/svn/trunk/doc/api/index.html}).

The collected tweets were filtered, first for Russian and Ukrainian language, identifying respective alphabet characters in the tweet text. Tweets, where Twitter users specified being from Russia or any region/town in Russia were removed from the data. We also filtered for SPAM tweets (around 17\% of all the tweets in our data), using the following keywords, based on word count analyses: \textsf{porn, phone, games, androidgames, minecraft, ipadgames, loosing weight, gold, mtvstars, crossword, barbie, sony, holidays, shop, TV series, volkswagen, sales, starlet, apartments, estate, bmw, mercedes, sex, happysex, viagra, stock, prostitutes, teen, price, diet, buy, credit} in English, Ukrainian and Russian language. Tweets containing these words were removed from the data. Furthermore, we filtered the data for political context with an extensive set of keywords to remove irrelevant tweets and therefore unnecessary noise from the data: \textsf{protest, mobilization, Ukraine, kiborgi, glory to the heroes (one word hashtag in Ukrainian and Russian - geroiamslava), junta, government, duma, kremlin, parliament, rada, premier ministry, president, minister, ministry, political action, demonstration, opposition, power, authorities, democracy, nationalists, communists, liberals, Putin, Poroshenko, political party, politics, politicians, political, policy, revolution, citizen, criticism, critics, agitation, sanctions, censoring, illegal, legal, solidarity, assembly, rally, law, regulation, resistance, civil disobedience, resist, reforms, communism, capitalism, administration, news, society, violation, RNBO, Iaijtheniukh, ukr, ukry, khokhly, okraina, ruina, raguli, poproshenko, papashenko, papasha, bacon to the heroes (a wordplay in Russian with glory to the heroes - geroiamsala), civil war, nazis, Bandera, banderlogi, bandery, banderovtsy, benderovets, visitka Iarosha, titushki, EuroMaidan, U\_revolution, peaceful march, radio freedom, inforesist, Aronets, avtomaijdan, Ukraine truth news (one word hashtag), SOS Maidan (one word hashtag), digital Maidan (one word hashtag), Maidan history (one word hashtag), sidemaidan, Maidan, freedom, Ukrainian, NATO for Ukraine (one word hashtag), will, choice, against, anti-Maidan, ukrop, Europe, USA, Kiev, Timoshenko, Cameron, dead, killed, negotiations, military service, Zakharchenko, police, Obama, war, activists, right sector, Azarov, upper, leader, Jatseniuk, Turchinov, Klitchko, MID, Nayyem, briginets, constitution, Avakov, MVD, Yushenko, Kuchma, Kravchuk, Kharkov, Grushevskogo, warriors, soldiers, prisoners, Euro, EU, military, Sloviansk, army, West, East, Mariupol, reconciliations, NATO, ATO, crimealook, conflict, Medvedchuk, Medvedchukov, geo-politics, confrontation, Strasbourg, crisis, Lutsenko, Tusk, peace, world, Vladimirova, Vladimir, relations, unity, national, help, oligarchs, glory to Ukraine (one word hashtag), hundred, over, own will, self-determination, power, annexion, annexing, separatism, separatist, rebels, freedom fighters, freedom fight, moskali (meaning Moscow sympathizers), [Russian] official, Putin khujlo (offensive word, one word hashtag), luganda, lungandon, daunbas, zrada, occupants, MGB, Zakhar, Russian peace (one word hashtag), Russian world (one word hashtag), Putin is murder (one word hashtag), Putler, Putin help (one word hashtag), putinism, insurgents, insurgency, small/so what Russia, Crimea is ours (one word hashtag), At least Crimea is our (one word hashtag), punitive, ptnpnkh, ptn, pnkh, vata, vatnik, glory to Russia (one word hashtag), Russia, provocation, Novorossiia, DNR, LNR, berkut, kiberberkut, glory, stopcrimeantatarsgenocide, russiainvadeukraine, russiaviolatedceasefire, stoprussianaggression, weapon, Russian, Medvedev, Moscow, legislation, Yanukovich, truth, Donetsk, Luhansk, Russian march (one word hashtag), Russians, Ukrainians, Non-russians, Navalny, monument, Lenin, western, eastern, geek, Lavrov, grad, humanitarian, anti-Russian, SSSR, Donbass, Bafana, elections, legislator, fire, Mirakova, ukraintsami, radio, Putin supporters (one word hashtag), Ukraine\_France, Zakharov, Mironov} in Ukrainian and Russian language. Tweets that contained at least one of these keywords were kept in the filtered data, otherwise removed.

Any analysis of Twitter data faces a number of well-known difficulties \citep{RuthsPfeffer14}. Some of them, e.g. the SPAM tweet problems, we have addressed already above. One potential problem is that the sample only includes public tweets from public Twitter accounts. This does not pose a problem in the context of our study though, since we are interested in the use of Twitter as an instrument of communication in the public sphere. Moreover, Twitter data is not representative, which again is rather unproblematic for our study because all political groups and their supporters are represented on Twittersphere, so the public debate on Twitter does overall mirror the general public debate and public opinions \citep{Stern14,Ronzhyn14}. Another potential issue is that the sample is based solely on the 1\% of all public tweets, which for instance makes it difficult to use the data as panel data (though we've done this to probe opinion changes within individuals, see Supplementary Fig. \ref{panel}, the number of observations is however drastically reduced). However, other Twitter data samples offered by Twitter (e.g. Gardenhose with 10\% of all pubic tweets) have to be purchased, which makes them often unaffordable for research purposes.

One aspect of Twitter data "richness" is that it is supposedly geo-referenced. However, as already mentioned in the main manuscript quite a large proportion of Twitter users do in fact not provide any geographical information. Frequently, even if "geo\_enabled" was activated by the Twitter user (thus the value was set to "true"), the actual "geo" or "place" "coordinates" value would be "null" because the device that was used to post the tweet had no geo-referencing (e.g. GPS) activated. The data that the user provides through the profile "location" tag is more often available, however, here we have to rely on the accuracy and honesty of the Twitter users. Moreover, if the information provided through "location" was imprecise, e.g. just "Ukraine", we could not use this information for plotting. Did we however have serious and specific information from the user, e.g. "Kiev", then we used that information to generate a longitude and a latitude coordinate value that we could then plot on a generated Ukraine map in Fig. 3C,D in the main manuscript. We deliberately ignored another potential geographical information, the "time\_zone", because it offers only imprecise geographical information. Overall, we could extract form the Twitter data of the last September week 2014 493 Twitter users for whom we had sufficiently precise geographical data. This data along with the calculated West and East scores were used to produce the Figures 3C and D in the main manuscript.

\subsubsection*{Twitter Data Analysis}
In order to understand the empirical polarization dynamic in Ukraine and furthermore in order to make the data usable for model calibration, it was necessary to identify the political opinion of the Twitter users in our data. First, we identified the Twitter users in our Twitter data based on the value of their "screen\_name". We then compiled two lists of keywords, extracted from the word count analysis of the most common words that contain words associated with either the pro-East political camp or the pro-West political camp: \textsf{annexion: -2 , annexing: -2, separatism: -2, moskali (meaning Moscow sympathizers): -2, [Russian] official: -2, Putin khuijlo (one word hashtag): -3, Luganda: -3, Lugandon: -3, Daunbas: -3, zrada: -2, insurgents: -2, occupants: -2, MGB: 1, Zakhar: 1, Russian peace (one word hashtag): -2, Putin is murder (one word hashtag): -3, Putler: -3, Putin help (one word hashtag): 2, putinism: -1, freedom fighters (one word): 2, freedom fight (one word): 2, small/so what Russia: 2, Crimea is ours (one word hashtag): 3,  At least Crimea is ours (one word hashtag): 3, punitive: 3, ptnpnkh: -3, ptn: -1, pnkh: -3, vata: -3, vatnik: -3, glory to Russia: 3, Russia: 0, Russian: 0, Russians: 0, provocation: 0, separatist: -3, separatism: -3, Novorossiia: 2, DNR: 1, LNR: 1, sanctions: 1, berkut: 1, kiberberkut: 1,  stopcrimeantatarsgenocide: -3, russiainvadedukraine: -3, russiaviolatedceasefire: -3, anti-Russian sanctions (one word hashtag): 2, stoprussianaggression: -3, peace: 0, Putin: 0, weapon: 0, Medvedev: 0, Crimea: 0, Moscow: 0, Yanukovich: 0, truth: 0, Donetsk: 0, from Lugansk (one word): 0, Russian march (one word hashtag): 0, Non-Russians: 0, dead/killed: 0, negotiations: 0, politce: 0, rebels: 0, Lugansk: 0, war: 0, duma: 0, Mockva: -1, government: 0, leader: 0, monument: 1, Lenin: 1, warriors: 0, prisonerns: 0, military: 0, East: 0, army: 0, relations: 0, soldier: 0, reconciliations: 0, geek: -2, Lavrov: 0, grad: -1, humanitarian: 1, anti-Russian: 2, SSSR: 1, Donbass: 0, Bafana: 0, elections: 1, referendum: 0, confrontation: 0, fire: 0, crisis: 0, conflict: 0, Putin supporters (one word hashtag): -1, Mironov: 0} were the keywords for the East political camp and \textsf{mobilization: 1, kiborgi: 2, unity: 1, glory to the heroes (one word hashtag): 3, RNBO: 1, Iaijtheniukh: -3, junta: -2, ukr: -3, ukry: -3, khokhly: -1, okraina: -1, Ruina: -3, raguli: -3, poproshenko: -2, Papashenko: -2, papasha: -3, bacon to the heroes (a wordplay in Russian with glory to the heroes - geroiamsala): -3, civil war (one word hashtag):- 2, visit Kaiarosha (one word hashtag): 2, nazis: -3, Bandera: -3, banderlogi: -3, bandery: -3, banderovtsy: -3, titushki: 2, Ukraine: 0, Euro Maidan (one word hashtag): 1, U\_REVOLUTION: 2, peaceful march (one word hashtag): 1, radio freedom (one word hashtag): 1, inforesist: 1,  Aronets: 1, avtomaijdan: 1, Ukrainian truth news (one word hashtag): 1, SOS Maidan (one word hashtag): 1, digital Maidan (one word hashtag): 1, Maidan history (one word hashtag): 1, sitemaidan:1, Maidan: 0, freedom:1, government: 1,  Ukrainian: 0, Urkrainians: 0,  NATO for Ukraine (one word hashtag): 2, will: 2, will: self-determination, against: 2, revolution: 2, anti-Maidan:- 2, ukrop: -3, Poroshenko: 0, Europe: 0, USA: 0, Kiev: 0, weapon: 0, Timoshenko: 0, truth: 0, Cameron: 0, democracy: 0, dead/killed: 0, negotiations: 0, military service: 0, Zakharchenko: 0, opposition: 0, Obama: 0, war: 0, parliament: 0, activists: 0, political action (one word): 0, protest: 0, sector: 0, Azarov: 0, upper: 0, leader: 0, rada: 0, Yatseniuk: 0, Turchinov: 0, Klichko: 0, MID: 0, Nayyem: 0, Briginets: 0, constitution: 0, president: 0, Avakov: 0, MVD: 0, Yushchenko: 0, Kuchma: 0, Kravchuk: 0, Kharkov: 0, Grushevskogo: 0,  warriors: 0, Euro: 0, EU: 0, military: 0, Slaviansk: 0, West: 0, western: 0, army: 0, soldier: 0, Mariupol: 0, reconciliations: 0, NATO: 0, ATO: 1, conflict: 0, Medvedchuk: 0, Medvedchukvv: 0, Strasbourg: 0, crisis: 0, Lutsenko: 0 ,Tusk: 0, peace: 0, relations: 0, national: 1, help: 1, glory to Ukraine (one word hashtag): 2, hundred: 2, over: 2, power: 2, BENDERovets: 3, civil war: -2, krimealook: 1, Ukraine\_France: 1} were the keywords for the West political camp. These keywords were scored between -3 and 3, with negative scores indicating negative attitudes towards the respective political camp, positive values positive attitudes and a zero a neutral opinion. Some keywords were unambiguously associated with a political affiliation. For instance, the word "Putler" (scored -3), a composition of Putin and Hitler, is clearly a negatively annotated word referencing the East political camp. Similarly, the word "banderovtsy" (scored -3) linking the West political camp to Stepan Bandera, leader of the Ukrainian nationalist and independence movement during the Second World War, who cooperated with Nazi Germany, shows clearly a strong disapproval of the West political camp. On the other hand the hashtag \#slavarossii (translated: Glory to Russia, scored +3), shows a clear support for the East political camp, while the hashtag \#natoforukraine (scored +3) expresses an unequivocal pro-Western political affiliation. A word like "Putin" (East political camp) or "Poroshenko" (West political camp) however would be assigned a "0" because depending on the remaining content of the tweet the names could have been associated with positive or negative attitudes. And for that reason an additional sentiment analysis of the tweets was necessary.

We used the sentiment analysis SentiStrength (\url{http://sentistrength.wlv.ac.uk}) approach to determine a sentiment score for each tweet. SentiStrength is a free Java-based automatic sentiment analysis tool, widely used in research, which is also available for the Russian language. We created a comprehensive dictionary for Ukrainian sentiment words (to be shared upon request) based on the SentiStrength sentiment scoring system and we reviewed the sentiment word dictionary that SentiStrength is using for the Russian language and complemented it with other sentiment words (among others offensive words that were missing in SentiStrength) to make it equivalent to the Ukrainian dictionary. The sentiment scores in SentiStrength and thus in our two sentiment words dictionaries range from -5 to +5 with 0 signifying a neutral word, negative values a negative sentiment and positive values a positive sentiment. The higher the absolute value the stronger the sentiment. For each tweet we calculated a sentiment score average based on the identified sentiment words in the tweet. We calculated a separate sentiment score for the East political camp related keywords and one for the West political camp related keywords. The emotional intensity score is averaging these two sentiment scores. 

In the case of unequivocal keyword tweets, the tweet score would derive from these keywords. For instance, if a keyword contained the words "Putler" (scored -3) and  the hashtag "stoprussianaggression" (scored -3), the tweet score would be the average of these two scores, thus -3.  Whenever a tweet was scored zero because of neutral keywords, we added to the score the respective sentiment score. Thus if an East political camp related tweet had a zero score (e.g. "Putin"), and the East political camp related sentiment score was -2 (for instance resulting from the sentiment word "zachvatil" (translated: grabbed)), then the tweet would get an East political affiliation score of -2. This is based on the assumption that users would express positive sentiments about terms associated with their own camp and/or negative sentiments towards terms associated with the other camp. Since most users would have posted several tweets, users were assigned a set of tweet scores, depending on the number of tweets and from these scores overall average scores, one East score, one West score, were calculated for each user, representing their political affiliation

Automatic classification and sentiment analysis have certainly their limitations, e.g. automatic sentiment analysis often fail to spot irony and classifications do not account for instance for complex inner-fraction dynamics, that is political fractions within political fractions. However, manual classification is becoming increasingly impossible with the growing amount of data and/or limited capacities and resources and thus automatic classification is increasingly applied. It is however important to work on further developing and elaborating these tools and to supervise, critically reflect and where required correct the process of automatic analysis and its outcomes.

\begin{figure*}
\begin{center}
\includegraphics[scale=0.34]{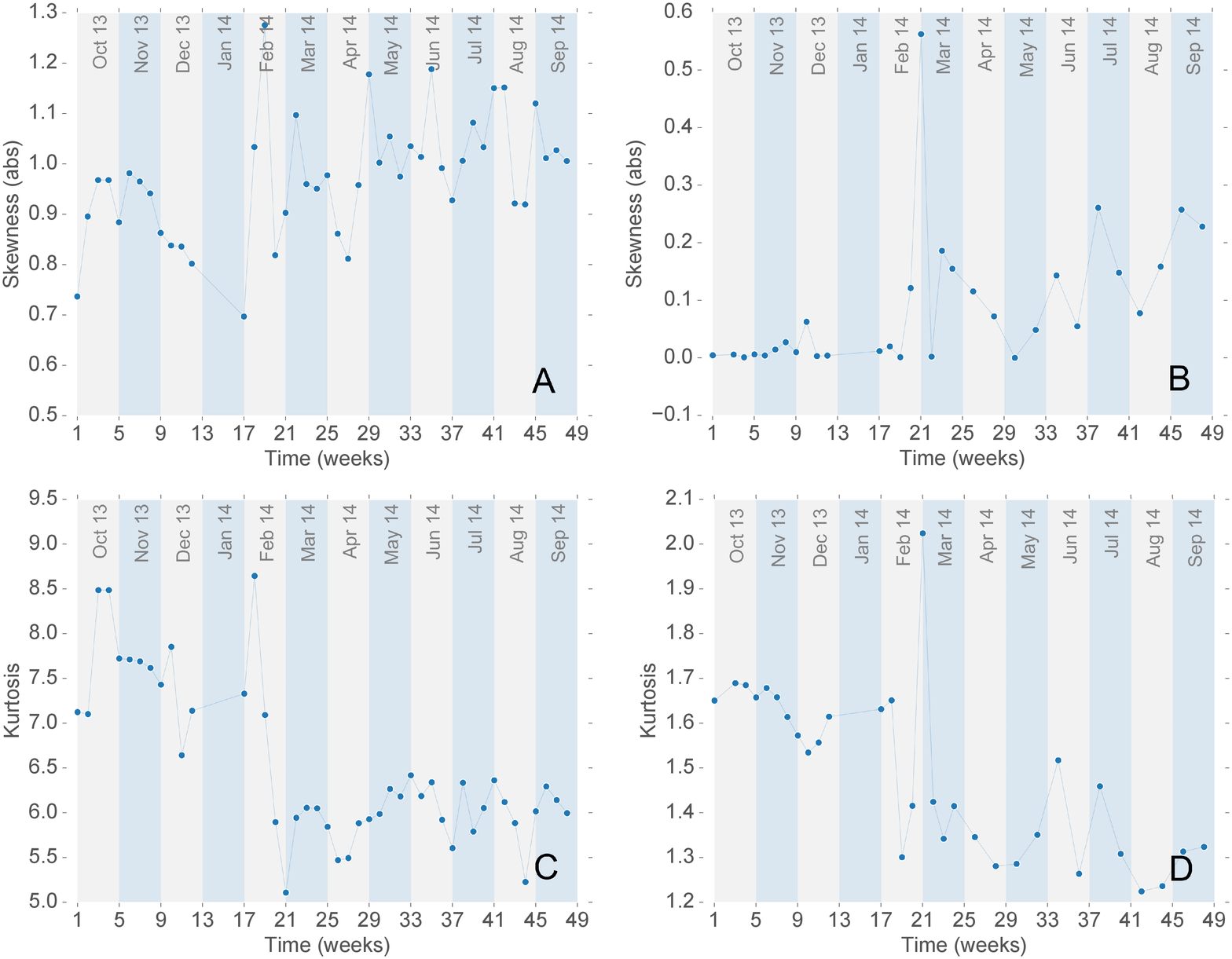}
\end{center}
\caption{(A) Time series plot of absolute skewness for opinion distributions in Twitter data. (B) Time series plot of absolute skewness for opinion distributions in simulations. (C) Kurtosis of the distribution of opinions in Twitter data. (D) Kurtosis of the distribution of opinions in simulations. Simulation parameters in plots B and D are $\alpha=0.15$ and $\eta=0.12$.}
\label{kurtosis_skew}
\end{figure*}

\section*{Additional Results}
We present here some additional results that support our main conclusions in the manuscript. Supplementary Figure \ref{kurtosis_skew} shows the skewness and kurtosis of opinions in the Twitter data and computer simulation. These plots show again the discontinuity of political opinions in the Ukrainian Twittersphere, with jumps in February 2014.


\begin{thebibliography}{10}

\bibitem{Belletal14}
James Bell, Katie Simmons, and Russ Oates.
\newblock {Despite concerns about governance, Ukrainians want to remain one
  country.} {Pew Research Center}.
\newblock \url{http://www.webcitation.org/6qqM20kvf}, 2014.
\newblock Accessed on: 30 May 2017.

\bibitem{EuropeanCommission16}
{European Commission. Humanitarian Air and Civil Protection.}
\newblock {Ukraine. ECHO Factsheet}.
\newblock \url{http://www.webcitation.org/70SmHLcRn}, 2016.
\newblock Accessed on: 30 May 2017.

\bibitem{BBC}
BBC.
\newblock Ukraine's sharp divisions.
\newblock \url{http://www.bbc.co.uk/news/world-europe-26387353}, 2014.
\newblock Accessed on: 30 May 2017.

\bibitem{cencus}
{Ukraine's 2001 census}.
\newblock \url{http://2001.ukrcensus.gov.ua}.
\newblock Accessed on: 14 Aug 2017.

\bibitem{HelbingBalietti11}
Dirk Helbing and Stefano Balietti.
\newblock How to do agent-based simulations in the future: From modeling social
  mechanisms to emergent phenomena and interactive systems design.
\newblock \url{http://www.santafe.edu/media/workingpapers/11-06-024.pdf}, 2011.
\newblock Accessed on: 30 May 2017.

\bibitem{BirkinMalleson12}
Mark Birkin and Nick Malleson.
\newblock Investigating the behaviour of {Twitter} users to construct an
  individual-level model of metropolitan dynamics.
\newblock \url{http://www.webcitation.org/70SmCQuDQ}, 2012.
\newblock Accessed on: 26 June 2018.

\bibitem{Serranoetal15}
Emilio Serrano, Carlos~A. Iglesias, and Mercedes Garijo.
\newblock A novel agent-based rumor spreading model in {Twitter}.
\newblock In {\em Proceedings of the International World Wide Web Conference
  (IW3C2)}, 2015.

\bibitem{Randetal15}
William Rand, Jeffrey Herrmann, Brandon Schein, and Ne\v{z}a Vodopivec.
\newblock An agent-based model of urgent diffusion in social media.
\newblock {\em Journal of Artificial Societies and Social Simulation},
  18(2):DOI: 10.18564/jasss.2616, 2015.

\bibitem{goncalvez.b:2011}
Bruno Gon\c~alves, Nicola Perra, and Alessandro Vespignani.
\newblock Modeling users' activity on {Twitter} networks: Validation of
  dunbar's number.
\newblock {\em PLOS ONE}, 6(8):1--5, 08 2011.

\bibitem{weng.l:2012}
L.~Weng, A.~Flammini, A.~Vespignani, and F.~Menczer.
\newblock Competition among memes in a world with limited attention.
\newblock {\em Scientific Reports}, 2:00335, 03 2012.

\bibitem{takayasu.m:2015}
Misako Takayasu, Kazuya Sato, Yukie Sano, Kenta Yamada, Wataru Miura, and
  Hideki Takayasu.
\newblock Rumor diffusion and convergence during the 3.11 earthquake: A
  {Twitter} case study.
\newblock {\em PLOS ONE}, 10(4):1--18, 04 2015.

\bibitem{Suhay15}
Elizabeth Suhay.
\newblock Explaining group influence: The role of identity and emotion in
  political conformity and polarization.
\newblock {\em Political Behavior}, 37(1):221--251, 2015.

\bibitem{Isenberg86}
Daniel~J. Isenberg.
\newblock Group polarization: A critical review and meta-analysis.
\newblock {\em Journal of Personality and Social Psychology}, 50(6):1141--1151,
  1986.

\bibitem{MoscoviciZavalloni69}
Serge Moscovici and Marisa Zavalloni.
\newblock The group as a polarizer of attitudes.
\newblock {\em Journal of Personality and Social Psychology}, 12(2):125--135,
  1969.

\bibitem{DeGroot74}
Morris~H. DeGroot.
\newblock Reaching a consensus.
\newblock {\em Journal of the American Statistical Association},
  69(345):118--121, 1974.

\bibitem{Sunstein02}
Cass~R. Sunstein.
\newblock The law of group polarization.
\newblock {\em The Journal of Political Philosophy}, 10(2):175--195, 2002.

\bibitem{BaronRoper76}
Robert~S. Baron and Gard Roper.
\newblock Reaffirmation of social comparison views of choice shifts: Averaging
  and extremity effects in an autokinetic situation.
\newblock {\em Journal of Personality and Social Psychology}, 33(5):521--530,
  1976.

\bibitem{Myersetak77}
David~G. Myers, Sandra~Brown Wojcicki, and Bobette~S. Aardema.
\newblock Attitude comparison: Is there ever a bandwagon effect?
\newblock {\em Journal of Applied Social Psychology}, 7(4):341--347, 1977.

\bibitem{McPhersonetal01}
Miller McPherson, Lynn Smith-Lovin, and James~M. Cook.
\newblock Birds of a feather: Homophily in social networks.
\newblock {\em Annual Review of Sociology}, 27:415--444, 2001.

\bibitem{Currarinietal09}
Sergio Currarini, Matthew~O. Jackson, and Paolo Pin.
\newblock An economic model of friendship: Homophily, minorities, and
  segregation.
\newblock {\em Econometrica}, 77(4):1003--1045, 2009.

\bibitem{Dandekaretal13}
Pranav Dandekar, Ashish Goel, and David~T. Lee.
\newblock Biased assimilation, homophily, and the dynamics of polarization.
\newblock {\em PNAS}, 110(15):5791--5796, 2013.

\bibitem{Burnstein82}
Eugene Bernstein.
\newblock Persuasion as argument processing.
\newblock In Hermann Brandstatter, James~H. Davis, and Gisela
  Stocker-Kreichgauer, editors, {\em Contemporary problems in group
  decision-making}, pages 103--124. NewYork: Academic Press, 1982.

\bibitem{EbbsenBowers74}
Ebbe~B. Ebbsen and Richard~J. Bowers.
\newblock Proportion of risky to conservative arguments in a group discussion
  and choice shifts.
\newblock {\em Journal of Personality and Social Psychology}, 29(3):316--327,
  1974.

\bibitem{VinokurBurnstein78}
Amiram Vinokur and Eugene Bernstein.
\newblock Novel argumentation and attitude change: The case of polarization
  following group discussion.
\newblock {\em European Journal of Social Psychology}, 8(3):335--348, 1978.

\bibitem{LordLepper79}
Charles~G. Lord, Lee Ross, and Mark~R. Lepper.
\newblock Biased assimilation and attitude polarization: The effects of prior
  theories on subsequently considered evidence.
\newblock {\em Journal of Personality and Social Psychology},
  37(11):2098--2109, 1979.

\bibitem{Milleretal93}
Arthur~G. Miller, John~W. McHoskey, Cynthia~M. Bane, and Timothy~G. Dowd.
\newblock The attitude polarization phenomenon: Role of response measure,
  attitude extremity, and behavioral consequences of reported attitude change.
\newblock {\em Journal of Personality and Social Psychology}, 64(4):561--574,
  1993.

\bibitem{Munro02}
Geoffrey~D. Munro, Peter~H. Ditto, Lisa~K. Lockhart, Angela Fagerlin, Mitchell
  Gready, and Elizabeth Peterson.
\newblock {Biased assimilation of sociopolitical arguments: Evaluating the 1996
  U.S. presidential debate}.
\newblock {\em Basic and Applied Social Psychology}, 24(1):15--26, 2002.

\bibitem{TaberLodge06}
Charles~S. Taber and Milton Lodge.
\newblock Motivated skepticism in the evaluation of political beliefs.
\newblock {\em American Journal of Political Science}, 50(3):755--769, 2006.

\bibitem{Baker05}
Wayne~E. Baker.
\newblock {\em America's Crisis of Values: Reality and Perception}.
\newblock Princeton, NJ: Princeton University Press, 2005.

\bibitem{Dimocketal14}
Michael Dimock, Carroll Doherty, Jocelyn Kiley, and Russ Oates.
\newblock {Political polarization in the American public.} {Pew Research
  Center}.
\newblock \url{http://www.webcitation.org/70SlevVb7}, 2014.
\newblock Accessed on: 26 June 2018.

\bibitem{BaldassarriBearman07}
Delia Baldassarri and Peter Bearman.
\newblock Dynamics of political polarization.
\newblock {\em American Sociological Review}, 72(5):784--811, 2007.

\bibitem{Castellanoetal09}
Claudio Castellano, Santo Fortunato, and Vittorio Loreto.
\newblock Statistical physics of social dynamics.
\newblock {\em Reviews of Modern Physics}, 81(2):591--646, 2009.

\bibitem{Masetal13}
Michael M\"{a}s, Andreas Flache, K\'{a}roly Tak\'{a}cs, and Karen~A. Jehn.
\newblock In the short term we divide, in the long term we unite: Demographic
  crisscrossing and the effects of faultilines on subgroup polarization.
\newblock {\em Organization Science}, 24(3):716--736, 2013.

\bibitem{Deffuantetal00}
Guillaume Deffuant, David Neau, Frederic Amblard, and G\'{e}rard Weisbuch.
\newblock Mixing beliefs among interacting agents.
\newblock {\em Advances in Complex Systems}, 3:87--98, 2000.

\bibitem{HegselmannKrause}
Rainer Hegselmann and Ulrich Krause.
\newblock Opinion dynamics and bounded confidence models, analysis, and
  simulation.
\newblock {\em Journal of Artificial Societies and Social Simulation}, 5(3):1,
  2002.

\bibitem{Fortunatoetal05}
Santo Fortunato, Vito Latora, Alessa~Ndro Pluchino, and Andrea Rapisarda.
\newblock Vector opinion dynamics in a bounded confidence consensus model.
\newblock {\em International Journal of Modern Physics C}, 16(10):1535, 2005.

\bibitem{Lorenz07}
Jan Lorenz.
\newblock Continuous opinion dynamics under bounded confidence: A survey.
\newblock {\em International Journal of Modern Physics C}, 18(12):1819, 2007.

\bibitem{Bakshyetal15}
Eytan Bakshy, Solomon Messing, and Lada Adamic.
\newblock {Exposure to ideologically diverse news and opinion on Facebook}.
\newblock {\em Science}, 348(6239):1130--1132, 2015.

\bibitem{Conoveretal11}
Michael~D. Conover, Jacob Ratkiewicz, Matthew Francisco, Bruno Goncalves,
  Filippo Menczer, and Alessandro Flammini.
\newblock Political polarization on twitter.
\newblock In {\em Fifth International AAAI Conference on Weblogs and Social
  Media}, pages 89--96, 2011.

\bibitem{YardiBoyd10}
Sarita Yardi and Danah Boyd.
\newblock Dynamic debates: An analysis of group polarization over time on
  twitter.
\newblock {\em Bulletin of Science, Technology \& Society}, 30(5):316--327,
  2010.

\bibitem{AsherBandeiraSpaiser}
Molly Asher, Cristina~Leston Bandeira, and Viktoria Spaiser.
\newblock {Assessing the effectiveness of e-petitioning through Twitter
  conversations}.
\newblock In {\em Political Studies Association Annual Meeting 2017}, 2017.

\bibitem{GruzdTsyganova14}
Anatoliy Gruzd and Ksenia Tsyganova.
\newblock {Politically polarized online groups and their social structures
  formed around the 2013--2014 crisis in Ukraine}.
\newblock {\em Internet, Politics, Policy 2014: Crowdsourcing for Politics and
  Policy.}, 2014.

\bibitem{Stern14}
David Stern.
\newblock {The Twitter war: Social media's role in Ukraine unrest}.
\newblock \url{http://www.webcitation.org/70Slybzp7}, 2014.
\newblock Accessed on: 26 June 2018.

\bibitem{Ronzhyn14}
Alexander Ronzhyn.
\newblock The use of facebook and {Twitter} during the 2013-2014 protests in
  ukraine.
\newblock In {\em Proceedings of the European Conference on Social Media: ECSM
  2014}, 2014.

\bibitem{Colleonietal14}
Elanor Colleoni, Alessandro Rozza, and Adam Arvidsson.
\newblock {Echo chamber or public sphere? Predicting political orientation and
  measuring political homophily in Twitter using Big Data}.
\newblock {\em Journal of Communication}, 64(2):317--332, 2014.

\bibitem{Barbera15}
Pablo Barber\'{a}.
\newblock {Birds of the same feather tweet together: Bayesian ideal point
  estimation using Twitter data}.
\newblock {\em Political Analysis}, 23(1):76--91, 2015.

\bibitem{archive}
{Archive Team: The Twitter Stream Grab}.
\newblock \url{https://archive.org/details/twitterstream}.
\newblock Accessed on: 30 May 2017.

\bibitem{Leeetal17}
Mi~Kyung Lee, Ho~Young Yoon, Marc Smith, Hye~Jin Park, and Han~Woo Park.
\newblock Mapping a twitter scholarly communication network: a case of the
  association of internet researchers' conference.
\newblock {\em Scientometrics}, 112(2):767--797, 2017.

\bibitem{vicsek.t:1995}
Tamas Vicsek, Andras Czir\'{o}k, Eshel Ben-Jacob, Inon Cohen, and Ofer Shochet.
\newblock Novel type of phase transition in a system of self-driven particles.
\newblock {\em Physical Review Letters}, 75:1226--1229, 1995.

\bibitem{romensky.m:2014}
Maksym Romensky, Vladimir Lobaskin, and Thomas Ihle.
\newblock Tricritical points in a vicsek model of self-propelled particles with
  bounded confidence.
\newblock {\em Physical Review E}, 90:063315, 2014.

\bibitem{applet}
{Bounded Confidence XY Model: Online Simulator}.
\newblock \url{http://maksymromensky.com/bounded_confidence_xy_model}.
\newblock Accessed on: 24 June 2018.

\bibitem{Sobkowicz12}
Pawel Sobkowicz.
\newblock Discrete model of opinion changes using knowledge and emotions as
  control variables.
\newblock {\em PLoS ONE}, 7(9):1--16, 09 2012.

\bibitem{Sobkowicz13}
Pawel Sobkowicz.
\newblock Minority persistence in agent based model using information and
  emotional arousal as control variables.
\newblock {\em The European Physical Journal B}, 86(7):335, 2013.

\bibitem{Spaiseretal}
Viktoria Spaiser, Thomas Chadefaux, Karsten Donnay, Fabian Russmann, and Dirk
  Helbing.
\newblock {Communication power struggles on social media: A case study of the
  2011-12 Russian protests}.
\newblock {\em Journal of Information Technology \& Politics}, 14(2):132--153,
  2017.

\bibitem{Gallup15}
Julie Ray.
\newblock Ukrainians disillusioned with leadership.
\newblock \url{http://www.webcitation.org/70Sm5d0YG}, 2015.
\newblock Accessed on: 30 May 2017.

\bibitem{Cottiero15}
Christina Cottiero, Katherine Kucharski, Evgenia Olimpieva, and Robert~W.
  Orttung.
\newblock War of words: the impact of russian state television on the russian
  internet.
\newblock {\em Nationalities Papers}, 43(4):533--555, 2015.

\bibitem{Aro2016}
Jessikka Aro.
\newblock The cyberspace war: propaganda and trolling as warfare tools.
\newblock {\em European View}, 15(1):121--132, 2016.

\end{thebibliography}

\begin{thebibliography}{5}
\expandafter\ifx\csname natexlab\endcsname\relax\def\natexlab#1{#1}\fi
\expandafter\ifx\csname bibnamefont\endcsname\relax
  \def\bibnamefont#1{#1}\fi
\expandafter\ifx\csname bibfnamefont\endcsname\relax
  \def\bibfnamefont#1{#1}\fi
\expandafter\ifx\csname citenamefont\endcsname\relax
  \def\citenamefont#1{#1}\fi
\expandafter\ifx\csname url\endcsname\relax
  \def\url#1{\texttt{#1}}\fi
\expandafter\ifx\csname urlprefix\endcsname\relax\def\urlprefix{URL }\fi
\providecommand{\bibinfo}[2]{#2}
\providecommand{\eprint}[2][]{\url{#2}}

\bibitem[{\citenamefont{Bird et~al.}(2009)\citenamefont{Bird, Klein, and
  Loper}}]{BirdKleinLoper09}
\bibinfo{author}{\bibfnamefont{S.}~\bibnamefont{Bird}},
  \bibinfo{author}{\bibfnamefont{E.}~\bibnamefont{Klein}}, \bibnamefont{and}
  \bibinfo{author}{\bibfnamefont{E.}~\bibnamefont{Loper}},
  \emph{\bibinfo{title}{{Natural Language Processing with Python}}}
  (\bibinfo{publisher}{Sebastopol: O'Reilly}, \bibinfo{year}{2009}).

\bibitem[{\citenamefont{Perkins}(2010)}]{Perkins10}
\bibinfo{author}{\bibfnamefont{J.}~\bibnamefont{Perkins}},
  \emph{\bibinfo{title}{{Python Text Processing with NLTK 2.0 Cookbook}}}
  (\bibinfo{publisher}{Birmingham: PACKT}, \bibinfo{year}{2010}).

\bibitem[{\citenamefont{Ruths and Pfeffer}(2014)}]{RuthsPfeffer14}
\bibinfo{author}{\bibfnamefont{D.}~\bibnamefont{Ruths}} \bibnamefont{and}
  \bibinfo{author}{\bibfnamefont{J.}~\bibnamefont{Pfeffer}},
  \bibinfo{journal}{Science} \textbf{\bibinfo{volume}{346}},
  \bibinfo{pages}{1063} (\bibinfo{year}{2014}).

\bibitem[{\citenamefont{Stern}(2014)}]{Stern14}
\bibinfo{author}{\bibfnamefont{D.}~\bibnamefont{Stern}},
  \emph{\bibinfo{title}{{The Twitter war: Social media's role in Ukraine
  unrest}}},
  \bibinfo{howpublished}{\url{http://news.nationalgeographic.com/news/2014/05/140510-ukraine-odessa-russia-kiev-twitter-world/}}
  (\bibinfo{year}{2014}), \bibinfo{note}{accessed on: 30 May 2017}.

\bibitem[{\citenamefont{Ronzhyn}(2014)}]{Ronzhyn14}
\bibinfo{author}{\bibfnamefont{A.}~\bibnamefont{Ronzhyn}}, in
  \emph{\bibinfo{booktitle}{Proceedings of the European Conference on Social
  Media: ECSM 2014}} (\bibinfo{year}{2014}).

\end{thebibliography}
\end{document}